\documentclass[journal]{IEEEtran}
\usepackage{graphicx,mathtools,mathrsfs,amssymb}
\usepackage{multirow}
\usepackage{txfonts}
\usepackage{stfloats}
\usepackage{booktabs}
\usepackage{bm,url}
\usepackage{multicol}
\usepackage[justification=centering]{caption}

\begin{document}
%

\title{Room geometry blind inference based on the localization of real sound source and first order reflections}

\author{Shan~Gao,~Xihong~Wu,~Tianshu~Qu
\thanks{Tianshu Qu is with the Key Laboratory on Machine Perception (Ministry of Education), Speech and Hearing Research Center, Peking University, Beijing 100871, China, e-mail: (qutianshu@pku.edu.cn).}}
\markboth{IEEE/ACM Transactions on Audio, Speech and Language Processing,~Vol.~XX, No.~XX, XX~XXXX}%
{Shell \MakeLowercase{\textit{\emph{et al.}}}: Bare Demo of IEEEtran.cls for IEEE Journals}
\maketitle

\begin{abstract}
The conventional room geometry blind inference techniques with acoustic signals are conducted based on the prior knowledge of the environment, such as the room impulse response (RIR) or the sound source position, which will limit its application under unknown scenarios. To solve this problem, we have proposed a room geometry reconstruction method in this paper by using the geometric relation between the direct signal and first-order reflections. In addition to the information of the compact microphone array itself, this method does not need any precognition of the environmental parameters. Besides, the learning-based DNN models are designed and used to improve the accuracy and integrity of the localization results of the direct source and first-order reflections. The direction of arrival (DOA) and time difference of arrival (TDOA) information of the direct and reflected signals are firstly estimated using the proposed DCNN and TD-CNN models, which have higher sensitivity and accuracy than the conventional methods. Then the position of the sound source is inferred by integrating the DOA, TDOA and array height using the proposed DNN model. After that, the positions of image sources and corresponding boundaries are derived based on the geometric relation. Experimental results of both simulations and real measurements verify the effectiveness and accuracy of the proposed techniques compared with the conventional methods under different reverberant environments.  
\end{abstract}

\section{INTRODUCTION}
Room geometry inference is a process of estimating the reflective boundaries, allowing the geometry of the room to be inferences \cite{1-dokmanic2013acoustic}, \cite{2-tervo20123d}, \cite{3-yu2020room}. Unlike visual-based techniques, audio-based room geometry inference methods are conducted based on the recorded acoustic signals and play an increasingly important role in numerous applications, such as sound source localization \cite{4-kataria2017hearing}, \cite{4-2ribeiro2010turning}, \cite{4-3ribeiro2010using}, signal enhancement \cite{5-dokmanic2015raking}, immersive audio and augmented reality (AR) \cite{5-1-Betlehem}. 

The traditional methods for inferencing the room geometry are based on the measurement of Room impulse response (RIR), whose temporal structure is subsequently analyzed \cite{6-markovic2013estimation}, \cite{7-antonacci2012inference}, \cite{8-kuster2004acoustic}. The RIR is composed of direct sound, early reflections and later reverberation. The time delay between the direct sound and early reflections is determined by the sound source and the receiver's locations and the geometry of the room. Therefore, it is possible to derive the boundary information from the measured RIRs. Dokmanic et al. demonstrated that the room can be reconstruected from a single RIR by using the information of direct sound and reflections up to second order \cite{5-dokmanic2015raking}. Considering that the second-order reflections are hard to obtain, some researchers achieve this target by using the mobile receiver or a large microphone array that uses the first-order reflection information in RIR measured at different positions to estimate the boundary parameters \cite{9-peng2015room}. However, the RIR-based room geometry estimation methods need to activate the sound source with specific signals, which can not be used in the scenarios such as the sound source is unknown. 

Unlike the above approaches based on the measurement of multiple room impulse responses, Mabanda et al. proposed a room geometry inference method by introducing the beamforming techniques for a compact spherical microphone array \cite{10-mabande2013room}. This kind of method does not need the prior knowledge of the source signals but only the position of the sound source, which makes it more widely used. Nevertheless, the room geometry is challenging to reference based on one-point measurement because of the limitation of the current DOA and TDOA estimation algorithms. Therefore, multiple measurements of the first-order reflections corresponding to different boundaries are required to estimate the room geometry thoroughly. 

With the development of machine learning algorithms, researchers have proposed many room geometry estimation techniques by using the deep neural network model. For example, Yu et al. realized the room acoustic parameters estimation from a single RIR using the proposed DNN model \cite{3-yu2020room}. Multiple RIRs are used to increase estimation precision by averaging estimates. Furthermore, Niles et al. investigated the performance of a convolutional-recurrent neural network for blind room geometry estimation, which is conducted based on higher-order Ambisonics (HOA) signals \cite{11-poschadel2021room}. Although this end-to-end method has proved to be effective under specific circumstances, it ignores the position information of the sound source and the receiver in the room environment, which also determine the spatial characteristics of reverberation signals. This setting will result in insufficient constraints in solving room geometry estimation problems and affect the network model's reliability and the accuracy of the results.

In this paper, we proposed a novel room geometry estimation method based on the localization of the real sound source and first-order image sources. This work is inspired by the related work proposed by Mabande et al. in Ref. \cite{10-mabande2013room}  and further improved, which can be embodied in the following points. First, we have realized the sound source localization based on the geometric relationship between the real and image source. Therefore, the location of the sound source does not need to be given in advance. Besides, the direction of arrival (DOA) and time difference of arrival (TDOA) of direct signal and reflections are estimated using the proposed DNN models, so the perception of boundary position can be realized through one-point measurement. In contrast to most approaches found in the literature, the proposed method does not involve measuring RIRs and any other prior knowledge of the source. Therefore, it can be applied to many applications, e.g., speech enhancement or immersive audio recording. 

This paper is organized as follows: an overview of the proposed room geometry blind inference method is given in Sec. \uppercase\expandafter{\romannumeral2}. The proposed DOA and TDOA estimation methods that are processed in the Eigen beam (EB) domain are briefly reviewed in Sec. \uppercase\expandafter{\romannumeral3}. Sec. \uppercase\expandafter{\romannumeral4} and Sec. \uppercase\expandafter{\romannumeral5} have analyzed the geometric relation between the direct source and first-order reflections, based on which the sound source localization methods are proposed, and then the room geometry is inferred. The experiments and results are described in Sec. \uppercase\expandafter{\romannumeral6}, which proves the effectiveness and robustness of our method, and the conclusions are given in Sec. \uppercase\expandafter{\romannumeral7}. 

\section{Room geometry estimation method}

In this section, a room geometry estimation method is proposed, which can be performed under enclosure rooms with the shape of a convex polyhedron. This work is conducted based on the assumption that the impinging waves are reflected from the boundaries in a specular fashion, which is justified for most room boundaries with a wide frequency range \cite{12-allen1979image}. In this case, the first-order image sources can be viewed as the real source's mirror point concerning the corresponding boundaries. Therefore, the estimation of room boundaries can be achieved by the localization of the real sound source and first-order image sources based on the recorded microphone array signals. The proposed method does not involve measuring RIRs and prior knowledge of the sound source. So, it can be applied in the process of the other sound source processing algorithms, such as speech enhancement or immersive audio recording. 

\begin{figure}[hbt]
\centering
\includegraphics[width=17pc]{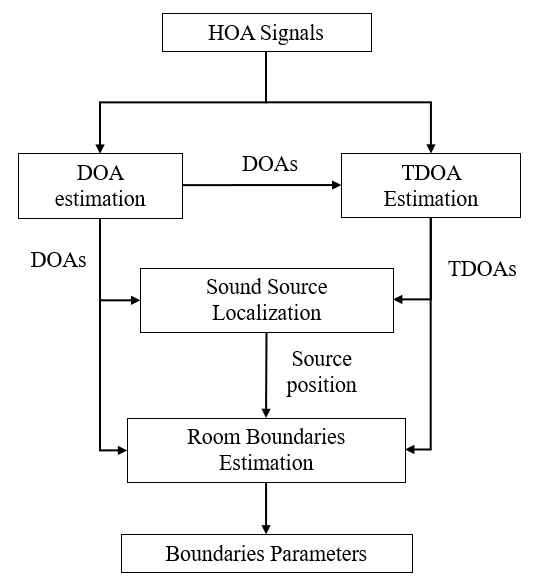}
\caption{The framework of the proposed method.}
\label{alg_framework}
\end{figure}

The procedure of the proposed techniques is expressed in Fig.\ref{alg_framework}. The DOA of the direct signal and reflections are firstly estimated, based on which the signals from the corresponding directions are then extracted, and the TDOAs between the direct signals and first-order reflections are estimated. After that, the DOA and TDOA information is combined to local the sound source and the image sources based on the geometric criterion. In the last, the desired geometric boundary parameters are inferred. 
Due to the absorption of the boundaries and the attenuation during propagation, the power of the reflected signals is lower than that of the direct signal. Therefore, signal extraction of first-order reflections is difficult because of their low signal to interference-noise ratio (SINR) and their coherence with the direct signals and among each other. To alleviate the disturbance of the no-target signals and background noise, the DNN-based methods for TDOA estimation and source localization are proposed in Sec. \uppercase\expandafter{\romannumeral3}

\section{DOA and TDOA estimation}

In our work, the reverberant signals are recorded using a spherical microphone array. And then, the recorded multiple-channel microphone array signals are transformed to the Eigen beam (EB) domain based on the spherical harmonics’ decomposition theory \cite{13-rafaely2004analysis}. So, the spatial sound field is represented by a set of orthogonal bases, which provides an elegant mathematical framework for spatial signal processing and facilitates the localization and extraction of target sources. 
To reduce noise amplification and suppress spatial aliasing error while retaining the spatial characteristic of of estimation EB domain signals under the reverberation environment, this process can be realized using the sparse DNN model \cite{14-gao2022sparse}. The obtained spherical harmonics signals can be expressed as

\begin{equation}
\bold{B}(kr) = [B_0^0(kr), B_1^{-1}(kr), B_1^0(kr), \ \cdots, B_N^N(kr)]^T,
\label{hoa signal}
\end{equation}
which is also regarded as the HOA signals. In the above equation, \(k\) is the wave number and \(r\) is the radius of the microphone array. \(B_n^m (\cdot)\) is the EB domain signal of order \(n\) degree \(m\), and \(N\) is the truncation order. The following DOA and TDOA estimation are all performed based on the HOA signals. 

\subsection{DOA estimation}

DOA estimation of room reflections is challenging due to the high coherence and low energy of the reflected signals related to the direct signal. To solve this problem, researchers have given a variety of solutions. The steered beamformer-based and subspace-based reflections localization techniques have been compared in Ref.\cite{15-sun2012localization}, and the results prove that the EB domain MVDR (EB-MVDR) is more suitable for the DOA estimation of the reflections. To further improve the accuracy of reflected signals localization results, a DNN-based DOA estimation model (DCNN) is conducted for the DOA estimation of the direct and first-order reflected signals. This model uses the correlation matrix of HOA signals as the input feature and the deconvolutional network to construct the spatial pseudo-spectrum \cite{16-gao2022localization}, which performs better than the EB-MVDR algorithm and previous DNN-based method. Therefore, the DCNN model is used here for DOA estimation and is briefly reviewed in the following.

Given the HOA signals in the time domain \(B_t\), the covariance matrix of broadband EB signals can be calculated as \(R_t=B_t B_t^H\). Using \(R_t\) as the input of the neural network model, the spatial pseudo-spectrum (SPS) map correlated with the directions in the spherical coordinate system can be reconstructed. The peaks of the SPS determine the DOAs \(\Omega_i,i=0,1,2,…,I\), where \(I+1\) is the total number of the estimated DOAs, and thus \(i=0\) denotes the direct path and \(i=1, …, I\) denotes the first-order reflections. To ensure this, the supervision of the DCNN model is the SPS map with the peaks corresponding to the direct path and first-order reflections, and the initial microphone array signals are processed with the weighted prediction error (WPE) algorithms to eliminate the late reverberation \cite{17-yoshioka2012generalization}.

\subsection{TDOA estimation}

Having estimated the DOAs of the direct path and first-order reflections, the signals from the localized directions are firstly extracted using the robustness beamforming techniques. Then a statistical analysis of cross-correlations is performed to estimate the TDOA between the direct signal and each reflection. Since the accurate estimation of the covariance matrix used in the MVDR algorithms is disturbed by the coherent signals, a fixed beamformer is used here for the signals’ extraction. According to the analysis in Ref. \cite{18-yan2010optimal} and Ref. \cite{19-rafaely2005phase}, the pure phase mode beamformer in the EB domain can obtain the maximum white noise gain (WNG) for the case of isotropic noise, which corresponds to the reverberation environment. Therefore, to realize the optimal signal extraction under reverberant environment, given the direction of the sound source \(\Omega_i\), the coefficients vector of the beamformer can be written as the 

\begin{equation}
\bold{w}(\Omega_i) = [Y_0^0(\Omega_i), Y_1^{-1}(\Omega_i), Y_1^0(\Omega_i), \ \cdots, Y_N^N(\Omega_i)]^T,
\label{weight function}
\end{equation}
where \(Y_n^m (\cdot)\) is the spherical harmonics function of order \(n\) and degree \(m\).

Due to the limitation of the main lobe width in the beamforming process, the extracted reflected signal often contains the direct signal, which will decrease the accuracy of the traditional generalized cross-correlation (GCC) algorithm for TDOA estimation. To alleviate the influence of background noise or interference signals, traditional GCC-based TDOA estimation methods are always conducted with specific weight coefficients, e.g., GCC with Phase Transform (GCC-PHAT) \cite{20-kwon2010analysis}. This section proposed a convolutional neural network for the time delay estimation (TD-CNN) of the reflect signals related to the direct signal. The motivation is as follows: the convolutional neural network can extract and compare the phase difference of the input dual-channel signals, based on which the time delay is estimated. Besides, the weight coefficients of different frequencies are learned using the supervision method, which will make the proposed model more robust under reverberant and noisy environments.

\begin{figure}[hbt]
\centering
\includegraphics[width=12pc]{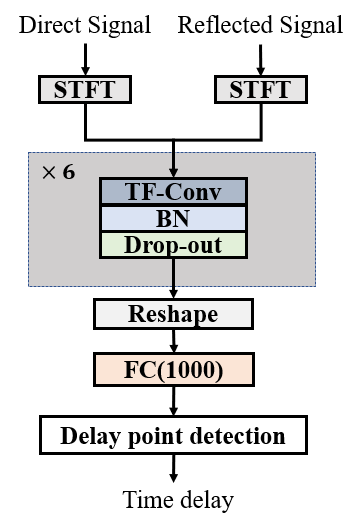}
\caption{Flow diagram of the TD-CNN model. TC-Conv denotes the convolutional layer followed by the batch normalization (BN) and drop out processing, and FC denotes the fully-connected layer for time delay classification with the nodes number.}
\label{td-cnn}
\end{figure}

We use the direct signal and reflect signal in the frequency domain as the input of the network. The real and imaginary parts of the dual-channel input signals are spliced and fed into a multiple-layer convolutional neural network to estimate the time delay. In order to ensure the accuracy and robustness of the network model, the length of the input signals should be much longer than the estimated time delay. Considering that the maximum delay of direct sound and early reflection under room reverberation environments is about 50ms \cite{1-dokmanic2013acoustic} and the signal sampling rate  \(f_s=16k\)Hz, we set the length of the network input signal as 5000 sampling points, that is, 313ms. The output layer contains 1000 nodes to represent the time delay with the exact time resolution as the reciprocal sampling rate. Ideally, only one node is activated, and its position represents the time delay. The enormous time delay obtained is about 62ms, which is sufficient for the TDOA estimation of the first-order signal.  Therefore, we formalize the time delay estimation problem into a classification problem, with the input signal \(x\in R^{5000\times2}\)and the output signal \(y\in R^{1000}\) The proposed neural network architecture is shown in Fig.\ref{td-cnn}, and the specific parameters are depicted in Table., which are set according to our pre-experiments. We use the cross between the network output and accuracy time delay representation as the loss function and use the Adam optimizer for the training of the network.

\newcommand{\tabincell}[2]{\begin{tabular}{@{}#1@{}}#2\end{tabular}} 
\begin{table}[htb]
\centering
\caption{Architecture of TD-CNN.}
\label{nn_arch}
\setlength{\tabcolsep}{2.1mm}{
\begin{tabular}{ccccc}
\toprule[2pt]
 \multirow{2}{*}{\tabincell{c}{Layer \\ name}}	  &  \multicolumn{3}{c}{Coefficients}  & \multirow{2}{*}{\tabincell{c}{Activation \\ function}} \\
 \cline{2-4}
 &Kernel size & strides & Output channels& \\
 \midrule[1pt]
 Conv1 & 1\(\times\)128  & 1\(\times\)5 & 128 & Leaky\_relu\\
 Conv2 & 1\(\times\)64   & 1\(\times\)5 & 256 & Leaky\_relu\\
 Conv3 & 1\(\times\)32   & 1\(\times\)5 & 256 & Leaky\_relu\\
 Conv4 & 1\(\times\)16   & 1\(\times\)3 & 512 & Leaky\_relu\\
 Conv5 & 1\(\times\)8    & 1\(\times\)3 & 256 & Leaky\_relu\\
 Conv6 & 1\(\times\)4    & 1\(\times\)3 & 128 & Leaky\_relu\\
 FC    &  \multicolumn{3}{c}{Output nodes: 1000} & Sigmoid 	\\
 \bottomrule[2pt]
\end{tabular}}
\end{table}

After the network is The TDOAs of the input dual-channel signals can be obtained by searching the maxima of the output layer
\begin{equation}
\tau = \lambda/f_s,
\label{weight function}
\end{equation}
where \(\lambda\) is the lag index of the maximum value, and \(f_s\) is the sampling rate. 

\section{Sound source localization}

Since the impact microphone array can hardly distinguish the wavefront curvature difference of the sources at different distances in the same direction, it is challenging to realize the sound source localization using such a microphone array. However, the existence of reverberant signals makes it possible since different sound source positions lead to different reverberation characteristics. The Gaussian mixture model or neural networks are conducted to realize sound source location. The input features of these models include amplitude or phase difference, binaural cross-correlation \cite{21-yiwere2017distance}, \cite{22-georganti2013sound}, direct sound reverberation energy ratio \cite{23-lu2010binaural} and others \cite{24-yiwere2019sound}. Considering that the mapping relationship between the sound source location and the above features is not fixed in different room environments, the current models trained in a specific room are hard to generalize in other rooms. 

In this section, we propose a novel sound source location method based on the first-order reflection of the sound source. Because the distance and orientation of the vertical walls are not fixed in different rooms, the corresponding reflection information is difficult to be directly used as a robust feature for estimating the sound source distance. However, the floor reflection always contains fixed characteristics, that is, the floor is generally horizontal and vertical to the microphone array, which can be used in the sound source localization process.

\begin{figure}[hbt]
\centering
\includegraphics[width=21pc]{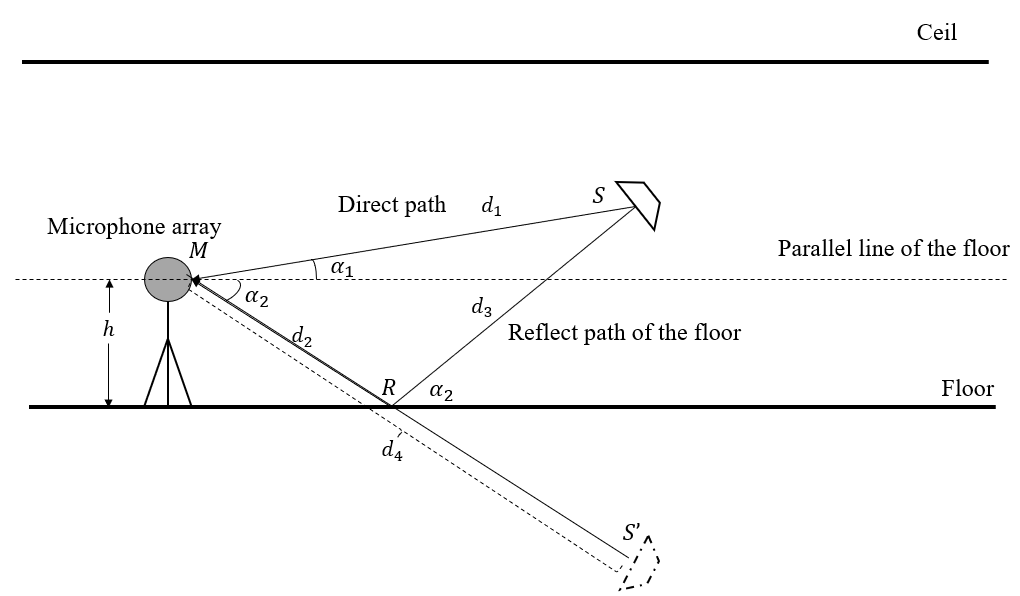}
\caption{The side view of the sound source and its reflection corresponding to the floor.}
\label{ref_pic}
\end{figure}

Fig.\ref{ref_pic} is the side view of the propagation mode of the sound source and its reflection corresponding to the floor, based on which the proposed sound source localization method is illustrated. The microphone array \(M\) is vertical to the floor with the height of \(h\), which is reasonable in applications and easy to obtain. The sound source signal propagates directly to the microphone array with the elevation \(\alpha_1\). At the same time, it also   the receiver through the reflection point \(R\) on the floor with the elevation \(\alpha_2\) and the corresponding image source is shown as \(S^{'}\). All of the angles can be estimated with the DCNN model. Then the included angle between MS and SR is \(\alpha_2-\alpha_1\), and the included angle between MR and SR is  \(\pi-2\alpha_2\), all of which can be derived. Using the above geometric information, the sound source distance is obtained according to the following clues.

\textbf{Sound source distance estimation based on array height (D-height)}

According to sine theorem, given the height of the microphone array, the sound source distance \(d_1\) can be estimated with the following formulation:
\begin{equation}
\frac{d_2}{sin(\alpha_2-\alpha_1)} = \frac{d_1}{sin(\pi-2\alpha_2)},
\label{crit 1}
\end{equation}
where \(d_2\) is the distance between the microphone array M and reflection point R, and can be calculated as \(d_2=h/sin(\alpha_2\)). Then the sound source distance can be estimated as
\begin{equation}
d_1=\frac{d_2sin(\pi-2\alpha_2)}{sin(\alpha_2-\alpha_1)} = \frac{h sin(\pi-2\alpha_2)}{sin(\alpha_2)sin(\pi-2\alpha_2)},
\label{driv 1}
\end{equation}

\textbf{Sound source distance estimation based on TDOA (D-TDOA)}

Given the TDOA \(\Delta T\) between the direct signal and the reflection from the floor, the distance from the microphone array to the image source S’ can be derived as 
\begin{equation}
d_4 = c\cdot\Delta T+d_1,
\label{driv 2}
\end{equation}

Since the projection distance between the image source and the real source is the same in the horizontal plane, the following equation can be established
\begin{equation}
d_4cos(\alpha_2) = d_1cos(\alpha_1),
\end{equation}
Then the sound source distance can be calculated as
\begin{equation}
d_1 = \frac{c\cdot\Delta T cos(\alpha_2)}{cos(\alpha_1-cos(\alpha_2)},
\label{driv 2}
\end{equation}

\textbf{Sound source distance estimation based on DNN (D-DNN)}

\begin{figure}[hbt]
\centering
\includegraphics[width=8pc]{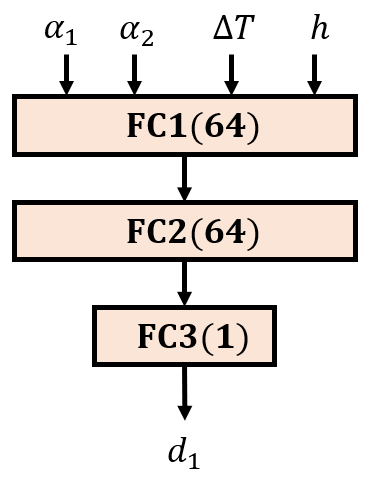}
\caption{The architecture of the sound source distance estimation DNN model. The coefficient in each fully connected layer denotes the nodes number. }
\label{dist-net}
\end{figure}

When the DOAs and TDOAs of the real and image source can be estimated, the sound source distance can be derived based on the above two criteria. 
However, due to the existence of the estimation error of above spatial clues, the use of a single feature is easily to cause a large sound source distance estimation error. In practice, effectively integrating different distance cues will reduce the impact of these estimation errors and improve the accuracy of the localization results. In order to achieve the above objectives, we construct a sound source distance estimation model based on a multi-layer feedforward network in this section. The height of the microphone array, the directions of the direct signal and floor reflection, as well as the TDOA information are used as the input of the model. We use three layers fully-connected neural network to extract the high dimensional information of the input feature. The sound source distance is obtained using the nonlinear character of the network with the regression method. The architecture and activation functions of the network are depicted in Fig.\ref{dist-net}. We use the mean square error between the network output and accuracy sound source distance as the loss function and use the Adam optimizer for the training of the network. In order to ensure the robustness of the proposed network model in applications, the zero-mean white Gaussian noise interference is added to the network input feature in the training process to simulate the input error that may occur in practice. The variance of the Gaussian noise is determined based on the trial method. 

\section{Room geometry estimation}

After obtaining the real sound source location information, the position of the boundaries can be estimated according to the image source direction and time delay information of each boundary to obtain the room geometry. Since the real sound source and its first-order image source are symmetrically distributed relative to the reflecting surface, the connecting line between the above two points can be viewed as the vertical line of the corresponding boundary. Therefore, the position of the vertical point and the direction of the vertical line can be used to uniquely determine the boundary, so as to realize the geometric reconstruction of the room. Here the center point of the microphone array is settled as the origin of the coordinate system. Assuming that the sound source position has been obtained, the DOA and TDOA of the first-order reflected signal of boundary \(j\) are expressed as \(\Omega_j\) and \(T_j\), respectively. The distance between the first-order image source \(S_j\) and the center point of the microphone array can be calculated by the following formula:
\begin{equation}
d_{S_j} = d_{S_0} + c\cdot T_j,
\end{equation}
where \(c\) is the sound speed, \(d_{S_0}\) is the distance between the real source and the microphone array. The location of image source \(S_j\) can be easily derived using \(d_{S_j}\) and \(\Omega_j\), which can be expressed as \((d_{S_j} cos(\Omega_j),\; d_{S_j} sin(\Omega_j))\). Having obtained the real source position \(\bold{P}_S\) and image source position \(\bold{P}_{\hat{S}_j}\), the position of the vertical point \(\bold{P}_{\hat{b}_j}\) and the direction of the vertical line \(\bold{n}_{\hat{b}_j}\) can be expressed as 
\begin{equation}
\bold{P}_{\hat{b}_j} = \frac{\bold{P}_S+\bold{P}_{\hat{S}_j}}{2},
\end{equation}
\begin{equation}
\bold{n}_{\hat{b}_j} = \frac{\bold{P}_S-\bold{P}_{\hat{S}_j}}
{||\bold{P}_S-\bold{P}_{\hat{S}_j}||},
\end{equation}

\section{EXPERIMENTS}

To verify the effectiveness and quantity the accuracy of the proposed methods, the network models are firstly trained based on the simulated data, and then, its performance with both simulations and real measurements is evaluated. Both speech and white Gaussian noise are used as the sound source signals in the test process, which ensure that the room modes across all frequencies are sufficiently excited. Note that in experiments with speech signals, the DOAs and TDOAs are only estimated during periods of speech activity. A simple energy-based voice activity detector (VAD) is used to ensure the accuracy of estimated results. 
The microphone array signals are simulated or recorded at a sampling rate of \(16\,k\)Hz with a frame length of 5000 samples and then processed offline. The angular difference between the look direction of two neighboring beams for DOA estimation and signal extraction is set to 3° both along azimuth and elevation. The effectiveness of the proposed TD-CNN model and sound source distance estimation method is firstly evaluated, and then the results of room geometry reconstruction are analyzed.

\subsection{Database}

For the training and testing of the proposed network, we create a simulation database under different room reverberant scenarios based on the image-source model \cite{12-allen1979image}. The length, width and height of rectangular rooms range from \(3\,m\times3\,m\times2\,m\) to \(10\,m\times10\,m\times4\,m\), and the reverberation time is randomly settled in the range from \(300\,ms\) to \(1000\,ms\). In each room, the reverberant signals are recorded using the spherical microphone array, which consists 32 microphones placed on a rigid sphere with a radius of \(4.2\,cm\), which has exactly the same geometry as the Eigenmike (EM32) \cite{25-EM32}. It can decompose the sound field for up to fourth-order spherical harmonics. Both the microphone array and the sound source are randomly distributed within the room with the minimum distance as \(0.5\,m\). The total number of rooms is 10000, corresponding to 10000 different simulated room impulse responses (RIRs), 80\(\%\) for training ,10\(\%\) for validation and 10\(\%\) for testing. The speech signals from the LibriSpeech database are used as the sources signals with the sampling rate as \(16\,k\)Hz. The length of the recorded signals in each room is about 10 seconds. Since the frame length is set to 5000 sampling point in the training and testing process, we have generated about 30 frames signals in each reverberant environment. 

\subsection{Evaluation metric}

The mean value \(T_{mean}\) and standard deviation \(T_{sd}\) of the time delay error with unit \(ms\) are used for evaluation of the proposed TD-CNN model. Although the direct signals and the first-order reflections are extracted based on the localization results, the time delay estimation might also fail because of the low SNR of extracted signals. Therefore, the accuracy estimation rate of the time delay \(R_{dect}\) is also used as the metric. To ensure the effectiveness of the model while reducing the impact of the interference on the results, a threshold value \(\sigma=0.3\), which is settled based on the experimental method. Only the maximum peak that exceeds \(\sigma\) is regarded as the valid estimation. Having obtain the DOAs and TDOAs of the direct and first order reflections, the  mean value \(S_{mean}\) and standard deviation \(S_{sd}\) of sound source distance error with unit \(m\) are calculated for the evaluation of the proposed sound source distance estimation method. Besides, for evaluating the accuracy of the estimated room boundaries, two parameters are calculated here \cite{10-mabande2013room}. The first parameter \(D_{S,\hat{S}_j}=D_{S}-\hat{D}_{S_j}\) is the difference between the distance \(\hat{D}_{S_j}\) from the estimated plane to the origin and the distance \(D_{S}\) from the 'ground truth' plane to the origin, while the second parameter \(\Theta_{\bm{n}_{j},\hat{\bm{n}}_j}\) is the angle between both planes' normal vectors.
\begin{equation}
\Theta_{\bm{n}_{j},\hat{\bm{n}}_j}=arccos(\bm{n}_{j}^T\cdot\hat{\bm{n}}_j),
\end{equation}
where \(\bm{n}_j\) and \(\hat{\bm{n}}_j\) is the normal vector of 'ground truth' plane and estimated plane.

\subsection{Results for simulated room}

\subsubsection{DOA and TDOA estimation}

We first conduct our experiments based on simulated data to evaluate the performance of the proposed methods under different scenarios. For insight into the DOA and TDOA estimation results, A specified simulated room with the dimension of
\(4\,m\times5\,m\times2.6\,m\) is used here. Set the lower-left corner of the room as the origin of the coordinate system, and the coordinates (unit: \(m\)) of the sound source and microphone array 
are (3.0, 3.0, 1.5)and (2.0, 2.0, 1.5), respectively. The reverberant time is set to \(800\,ms\). 

\begin{figure}[hbt]
\centering
\includegraphics[width=21pc]{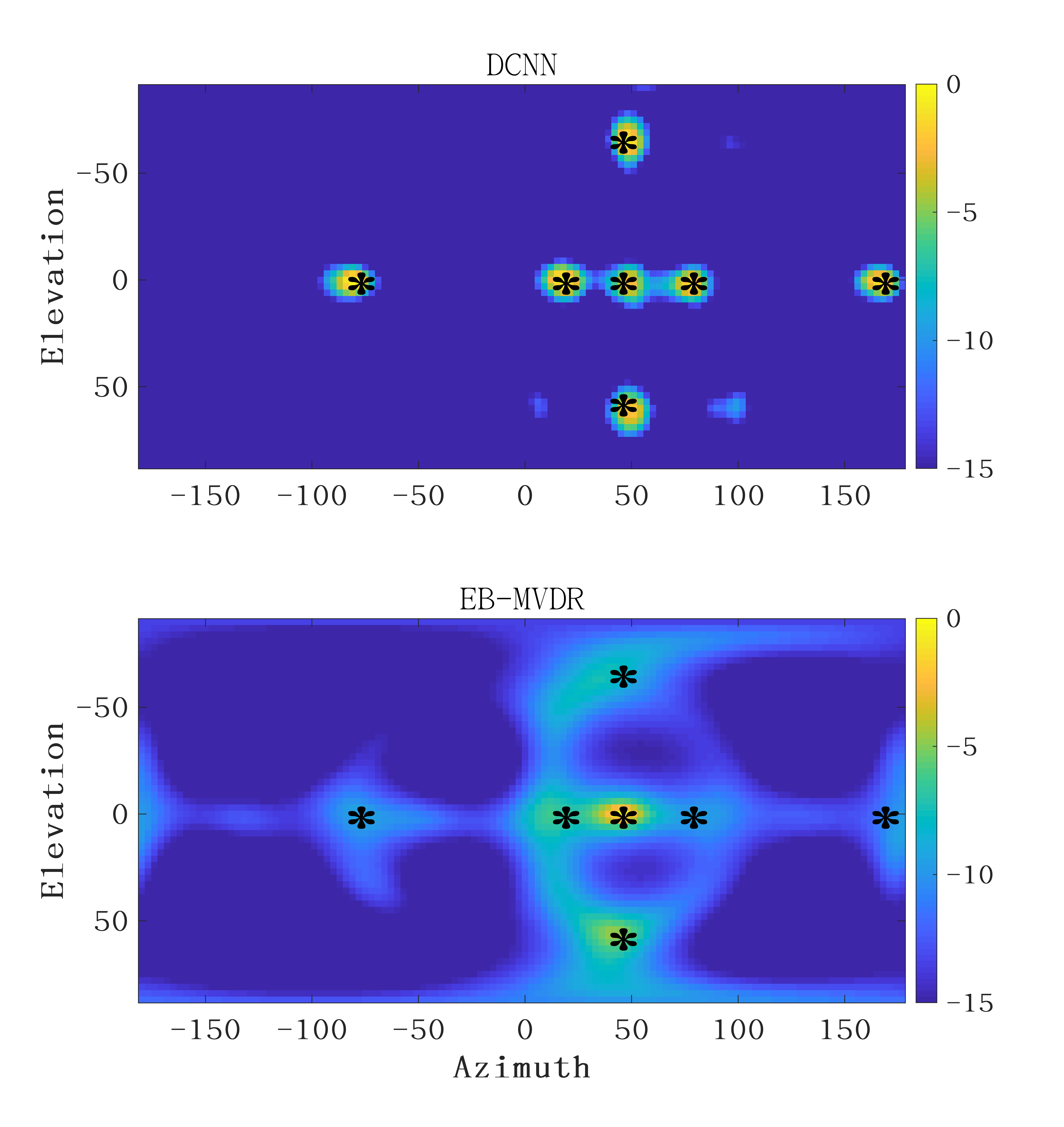}
\caption{The estimated spatial pseudo-spectrums of the DCNN model and EB-MVDR with frequency smoothing.}
\label{simu_doa}
\end{figure}

The exemplary spatial pseudo-spectrums (SPS) for DOA estimation are depicted in Fig.\ref{simu_doa}. Out of all peaks found in the SPS, only those that exceed the threshold \(\beta=-3\,dB\) are selected as the direct and first-order reflected signals’ directions. The value of \(\beta\) is determined experimentally to be a good compromise as to provide directions of direct signal and all first-order reflections while suppressing the disturbance of higher-order reflections and background noise. In all sub-figures depicting acoustic SPS, the ground truth DOAs for the direct source and first-order reflections are denoted with asterisks. Compared with the result of the MVDR method conducted in the Eigen beam domain with frequency smoothing (EB-MVDR), the results of the DCNN model has higher resolution and has the ability to obtain all first-order reflection directions by single point observation. 

\begin{figure}[hbt]
\centering
\includegraphics[width=21pc]{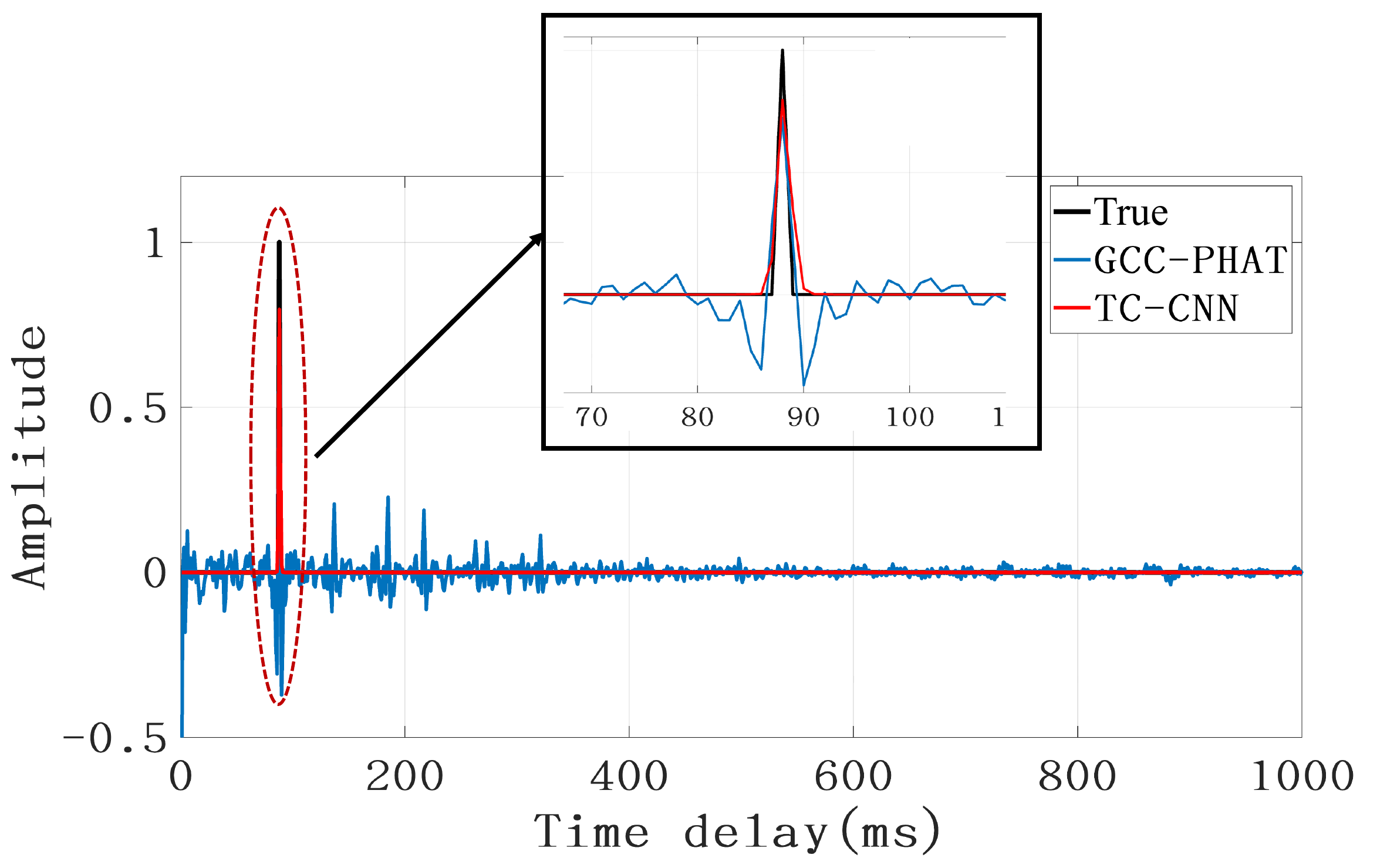}
\caption{The time delay estimation results of GCC-PHAT and TD-CNN model, and the black line denotes the real time delay. }
\label{simu-tdoa}
\end{figure}

\begin{figure}[hbt]
\centering
\includegraphics[width=20pc]{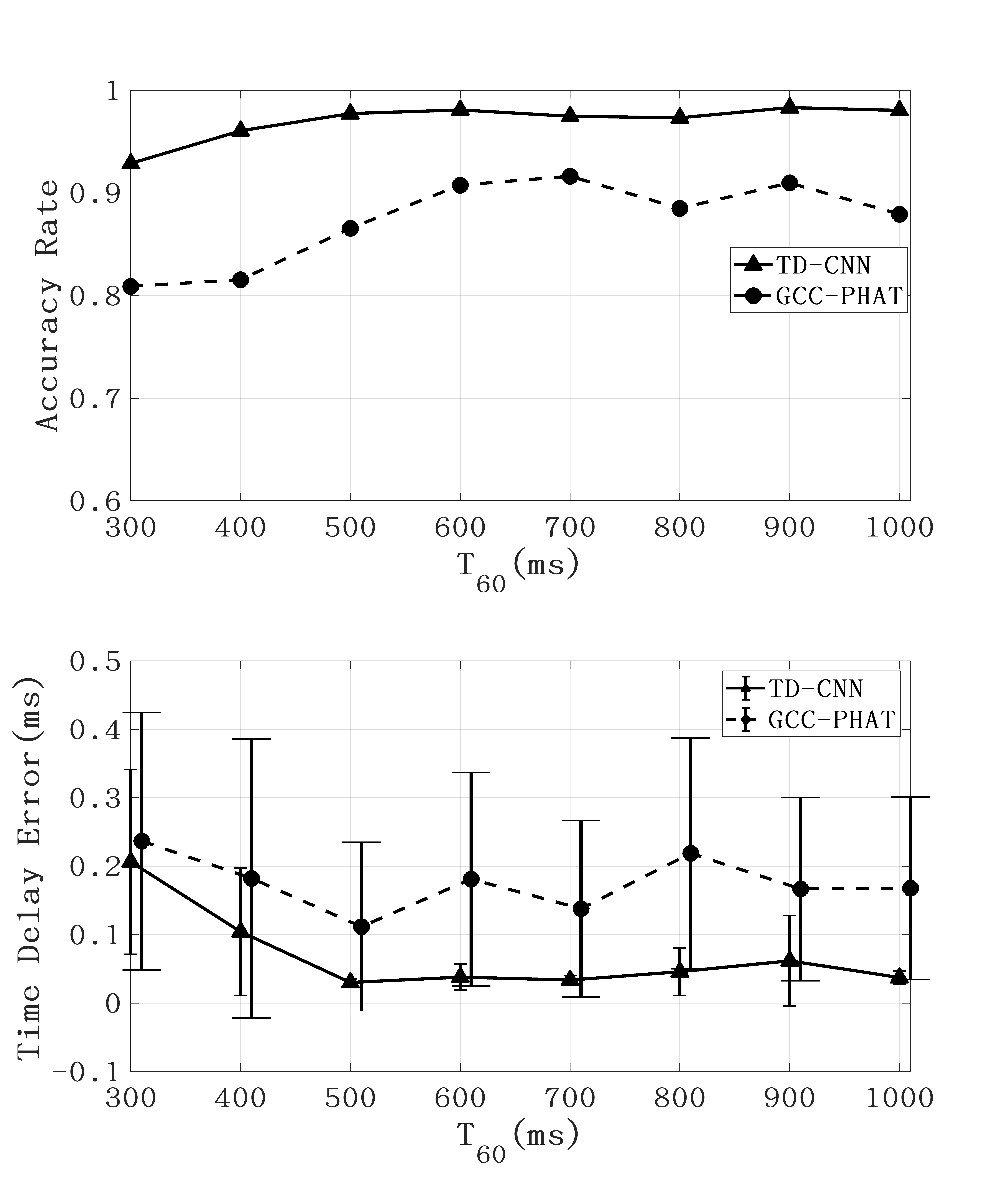}
\caption{Statistical result on the TDOA results under different rooms with different \(T_{60}\).}
\label{tdoa-1s}
\end{figure}

The estimated DOAs are used for signals extraction and TDOAs estimation. The time delay estimation results between the extracted direct signals and floor reflection are depicted in Fig.\ref{simu-tdoa}. Through both the results of GCC-PHAT and TD-CNN model present peaks at the right time delay position, the former contains several distinct peaks corresponding to the TDOAs of the direct signal and early reflections. In addition to the peak corresponding to the correct time delay, the interference of direct or non-target reflected signals will also lead to the peaks at the wrong position, which will confuse the estimation. In contrast, the proposed TD-CNN can effectively suppress the disturbance and makes the target time delay value more significant. To ensure the effectiveness of the model while reducing the impact of the interference on the results, a threshold value \(\alpha=0.3\), which is set based on the experimental method. Only the maximum peak that exceeds \(\alpha\) is regraded as the valid estimation. Besides, the peaks around the zeros-delay point (within 10 sampling points) is ignored to eliminate the influence of direct signal in the beamforming result of the reflected signals.

To verify the effectiveness of the proposed model under different reverberant scenarios, we make statistics on the TDOA results under different rooms with different \(T_{60}\), as shown in Fig.\ref{tdoa-1s}. In order to improve the results' differential of different algorithms, we have added different displacements to the error curves on the horizontal axis. Multiple groups of results near the same abscissa correspond to the same test environment, and the same processing is added to the following figures. It can be seen that the TD-CNN model has higher accuracy, as well as more minor angle error compared with the GCC-PHAT algorithm, which proves the robustness and effectiveness of the network model under different scenarios.

\subsubsection{Sound source Localization}

\begin{figure}[hbt]
\centering
\includegraphics[width=20pc]{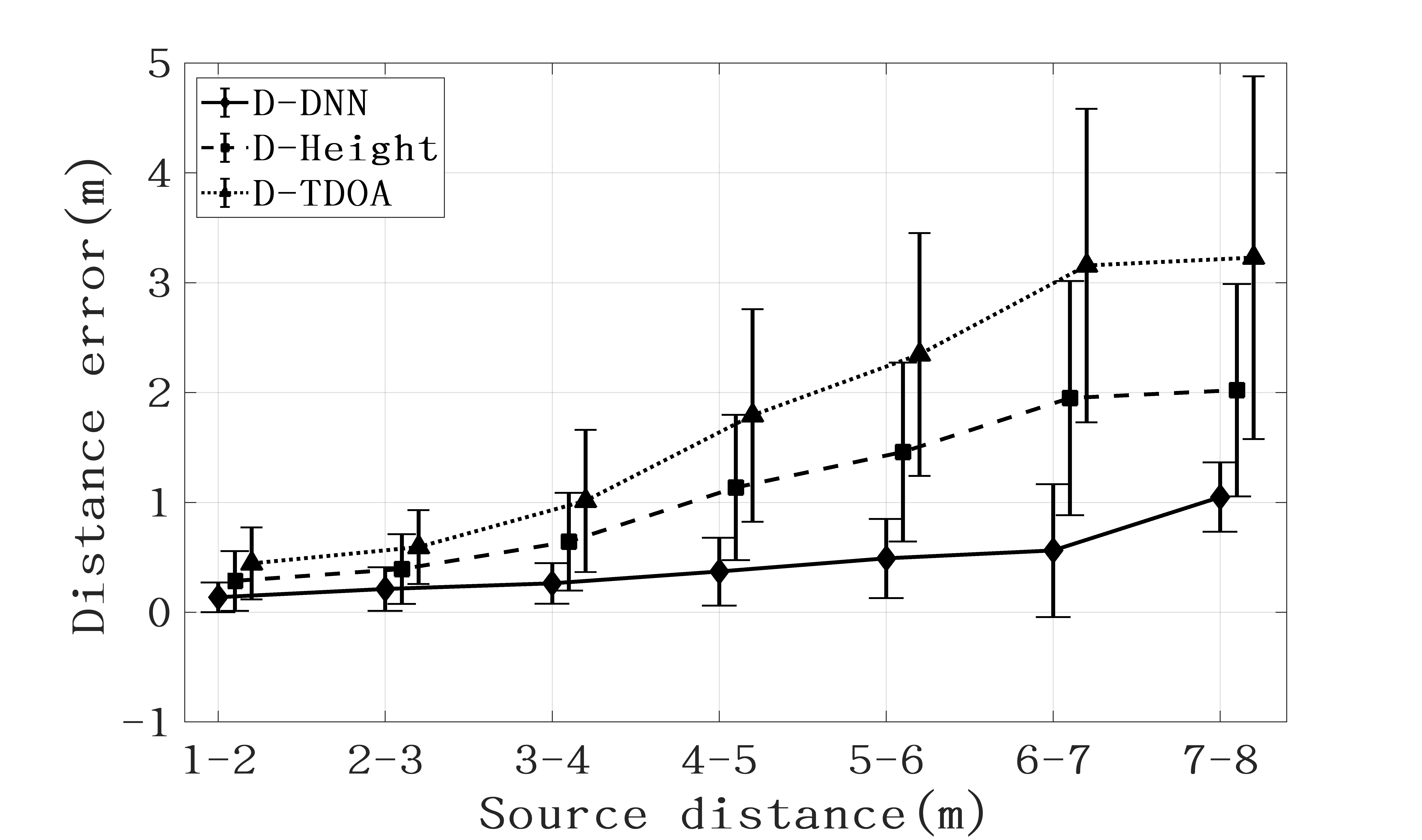}
\caption{The distance estimation error of different methods varies with the distance of sound source. (\(T_{60}=800ms\))}
\label{dist_d}
\end{figure}
\begin{figure}[hbt]
\centering
\includegraphics[width=20pc]{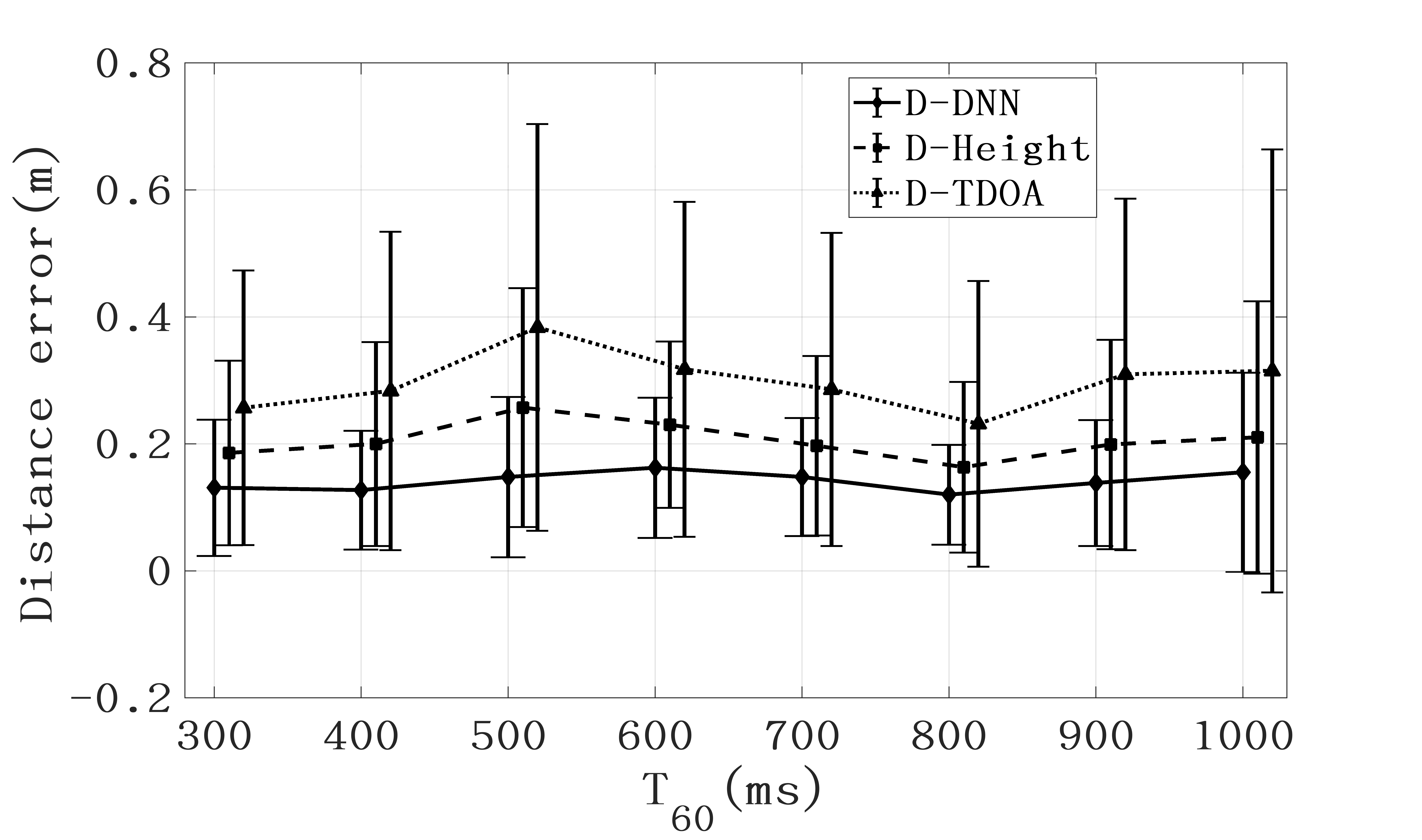}
\caption{The distance estimation error of different methods varies with \(T_{60}\), and the sound source distance range from 1\(m\) to \(2m\).}
\label{dist_r}
\end{figure}

Having obtained the DOAs and TDOAs of the direct and reflected signals, the sound source distance can be estimated based on the geometric derivations or the proposed neural network model. To evaluate the accuracy of these methods under different situations, we make statistics on the variation trend of the localization error with the distance of the sound source and the reverberation time \(T_{60}\), as shown in Fig.\ref{dist_d} and Fig.\ref{dist_r}. From the overall results, we can see that both the reflected sound delay information and array height information can be used as effective sound source localization features under different reverberant environments. Compared with the ground reflection delay information, sound source localization based on the height information of the array itself is much more accurate because there is no measurement error. The method based on the network model can effectively integrate the above features and obtain the optimal localization results under different cases. When the sound source distance is within \(8\,m\), the distance determination error can be basically maintained within 1m. 

\subsubsection{Room geometry inference}

\begin{figure}[hbt]
\centering
\includegraphics[width=20pc]{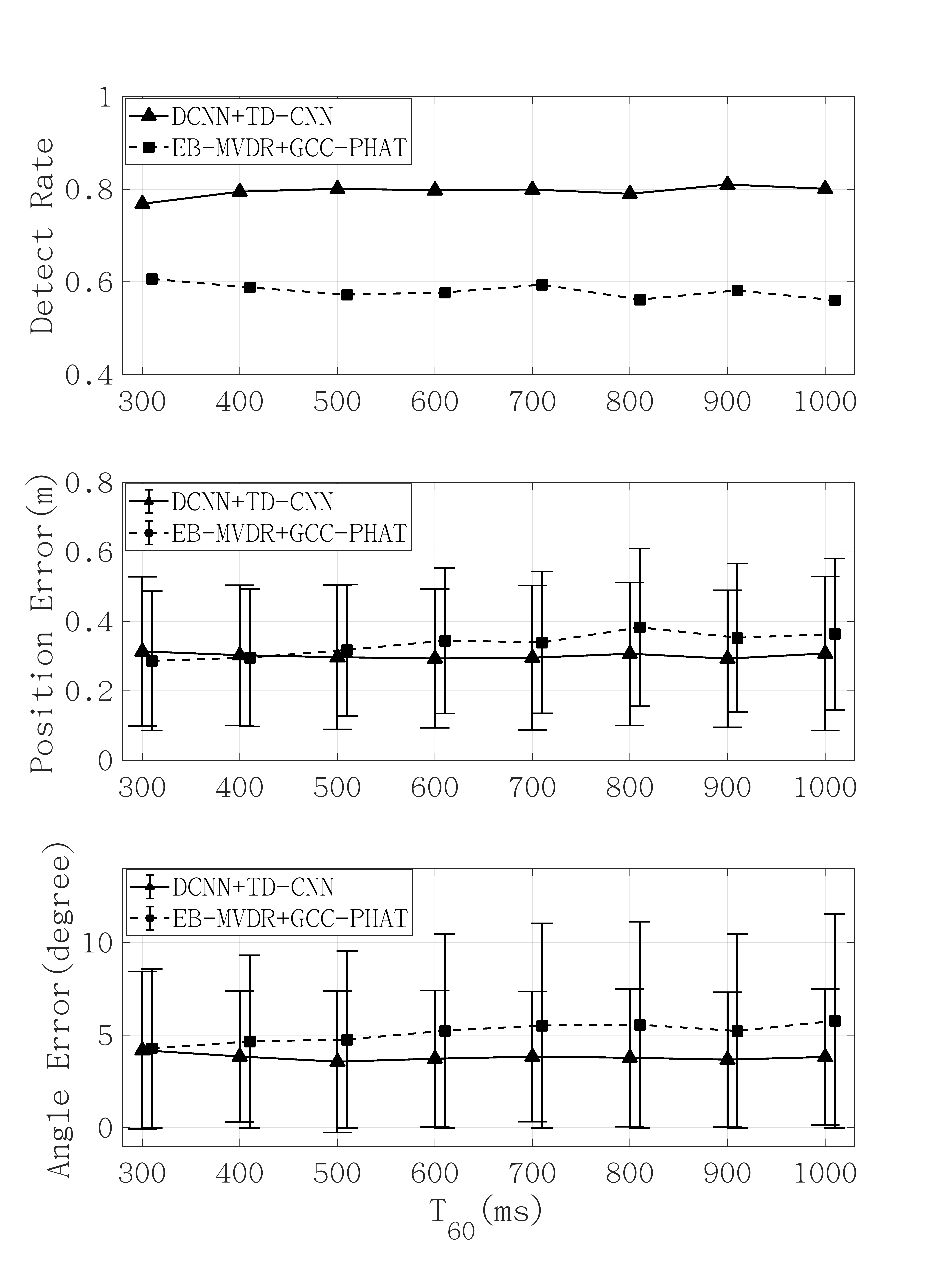}
\caption{The proposed neural network.}
\label{bound_error}
\end{figure}

Given the estimated sound source location, the DCNN and TD-CNN models are used for DOA and TDOA estimation of the reflected signals corresponding to each boundary, and the EB-MVDR and GCC-PHAT algorithms are presented as the baseline. In this process, we have trained a specific sound source distance estimation DNN model based on the estimated TDOA and DOA values from conventional methods. The detection rate and estimation errors of room boundaries under different tested reverberant scenarios are depicted in Fig.\ref{bound_error}.

From the results of the traditional method we can see that with the increasing of the reverberation time, the position and angle estimation error of the boundaries also increase gradually. Compared with the baseline system, the proposed method can effectively increase the sensitivity and accuracy of the boundaries detection, and is relatively robust in most reverberation environments. When \(T_{60}=800\,ms\), the average boundary detection rate in different environments is about 0.85, which means that the complete observation of room boundary can be realized in most cases. The exception may be caused by the excessive coincidence of the directions between direct and reflected signal. In these cases, the spatial resolution of DOA algorithms and the beamformer does not suffice to discriminate them well. This problem can be solved by moving the microphone array to observe the sound field from different angles. It should be noted that the position error of the proposed method is slightly higher than the traditional method when \(T_{60}=300\,ms\), which might be caused by the high detection rate of the reflections. The proposed method can derive more boundaries corresponding to the reflected signals with weak energy, which will bring much high localization and time delay estimation error. 

\subsection{Results for measured room}
\subsubsection{Experimental setting}

\begin{figure}[hbt]
\centering
\includegraphics[width=20pc]{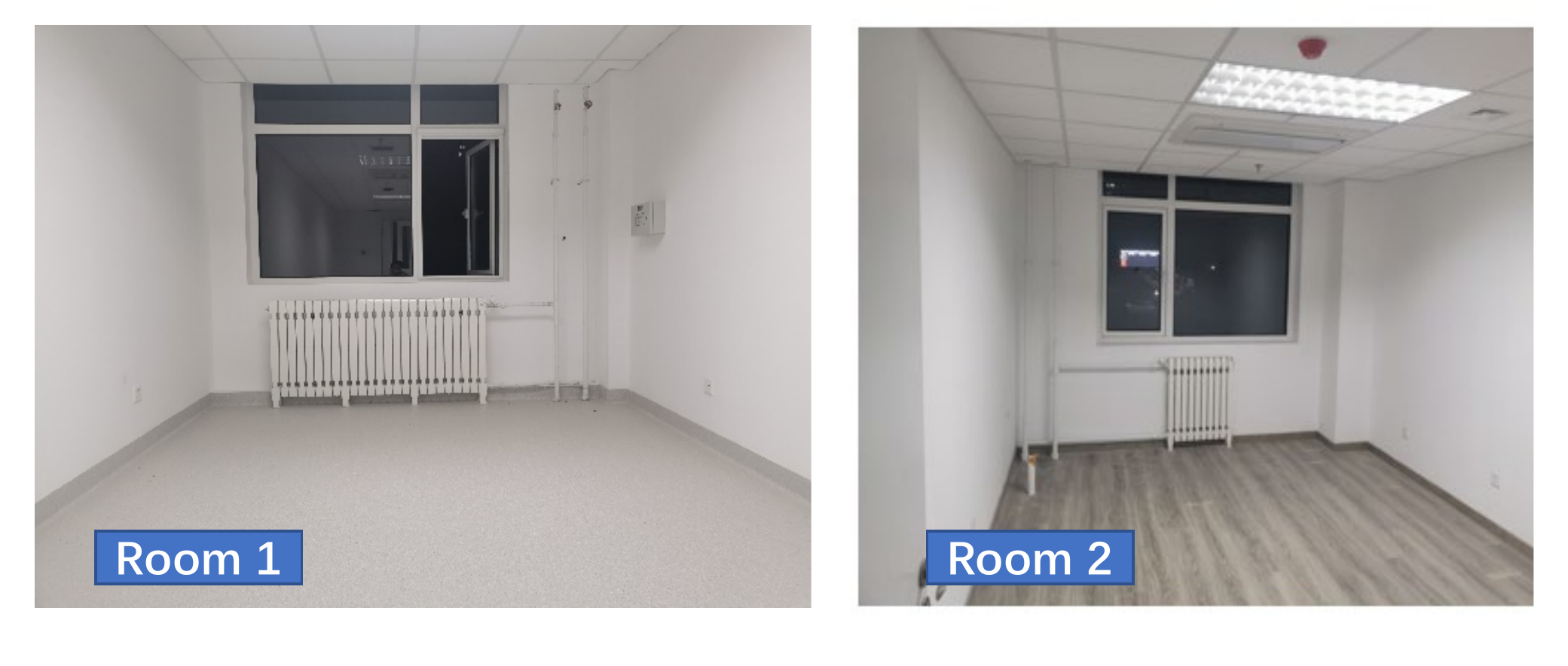}
\caption{The picture of measurement rooms.}
\label{room-fig}
\end{figure}

The measured signals are recorded using Eigenmike spherical microphone array under different rectangular rooms, as shown in Fig.\ref{room-fig}. The size of the rooms, as well as the location of the microphone array and the loudspeaker are shown in Table.\ref{room geo}. Each room has four walls with a smooth lime surface, one of which contains glass windows, the floor is a smooth stone floor (room 1) or a wood floor (room 2), and the ceiling is a porous gypsum board. Note that the inference results for the real room were compared to the “ground truth” obtained through manual measurements of the respective distances, and thus can be considered accurate up to manual measurement errors. The recording is performed at \(16k\)Hz sampling rate.

\begin{table}[htb]
\renewcommand\arraystretch{1.4}
\centering
\caption{Geometric parameters of the measurement rooms}
\label{room geo}
\setlength{\tabcolsep}{2.5mm}{
\begin{tabular}{ccc}
\toprule[2pt]
 Room Index & 1 & 2 \\
 \midrule[1pt]
 Room size\,(\(m\))& 6.47\(\times\)4.68\(\times\)2.34   & 5.20\(\times\)3.38\(\times\)2.34 \\
 Source position\, (\(m\)) & (4.27,3.21,1.40) & (3.47,2.09,1.25) \\
 Array position\,  (\(m\)) & (2.86,1.80,1.40) & (1.61,1.22,1.35) \\
 \bottomrule[2pt]
\end{tabular}}
\end{table}

\subsubsection{DOAs and TDOAs estimation}

\begin{figure}[hbt]
\centering
\includegraphics[width=20pc]{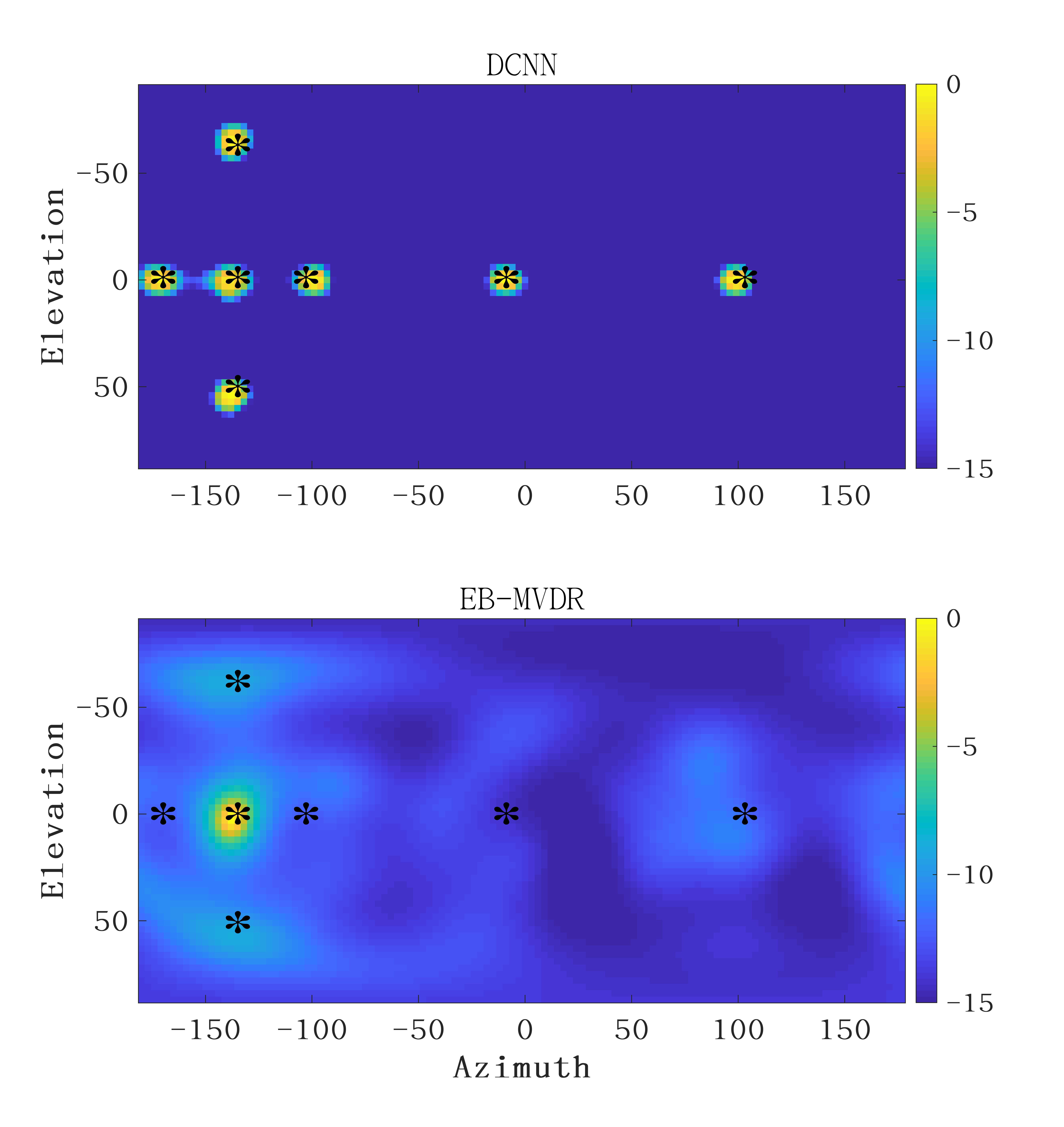}
\caption{The estimated SPS of DCNN and EB-MVDR methods using measurement data.}
\label{meas_doa}
\end{figure}

\begin{figure}[hbt]
\centering
\includegraphics[width=20pc]{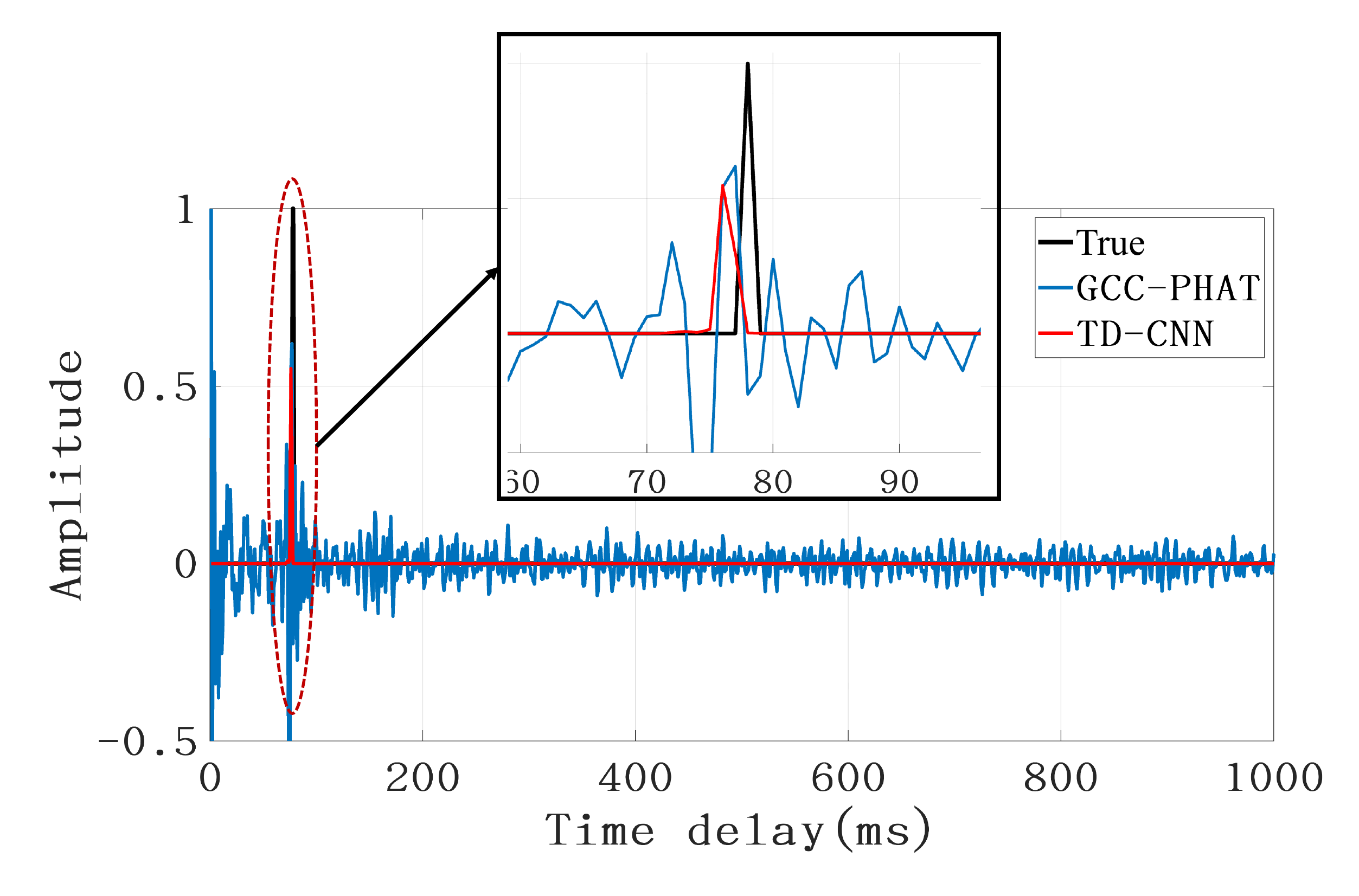}
\caption{The time delay estimation results of GCC-PHAT and TD-CNN model using measurement data.}
\label{meas_tdoa}
\end{figure}

Fig.\ref{meas_doa} and Fig.\ref{meas_tdoa} depict the estimated acoustic SPS and time delay information of room 2. In Fig.\ref{meas_doa}, seven peaks can be obviously found from the output of DCNN, corresponding to the directions of the direct signals and first-order reflections. It proves that the DCNN model has excellent generalization performance, and a better spatial resolution than the EB-MVDR algorithm in the actual environments. From Fig.\ref{meas_tdoa} we can see that both traditional methods and network models can achieve an approximate estimation of peak position. However, due to the interference of direct sound in the beamforming process of the reflected signal, the GCC-PHAT algorithm also has a significant peak near the delay of 0, which will cause the confuse in TDOA estimation. The TD-CNN output results can clearly indicate the delay information and suppress the interference of non-target signals. 

Table.\ref{tdoa error} shows the statistical results of delay estimation error and accuracy rate of the measurement signals. The average detection rate and mean square error of the TD-CNN model for related signals are 0.86 and 1.47\(ms\), which are significantly improved compared with the 0.79 detection rate and 2.01 \(ms\) of the GCC-PHAT method. Compared with the simulation results, the error of the measurement data increases significantly, which is mainly caused by the measurement errors of room size, microphone array, and source position. 

\begin{table}[htb]
\renewcommand\arraystretch{1.25}
\centering
\caption{Statistical results of time delay estimation methods using the measurement data}
\label{tdoa error}
\setlength{\tabcolsep}{2.5mm}{
\begin{tabular}{ccc}
\toprule[2pt]
 Method & GCC-PHAT & TD-CNN \\
 \midrule[1pt]
 \(R_{dect}\)     & 0.79 & \textbf{0.86} \\
 \(T_{mean}\,(ms)\) & 2.01 & \textbf{1.47} \\
 \(T_{sd}\,(ms)\)   & 4.16 & \textbf{2.93} \\
 \bottomrule[2pt]
\end{tabular}}
\end{table}

\subsubsection{Sound source localization}

Table.\ref{dist error} depicted the sound source distance estimation results using the above three methods. It can be seen that our method is able to estimate the distance of the source at about \(2\,m\) with an accuracy of about \(33\,cm\). To the best of the author’s knowledge, the best sound source distance estimation method based on DNN model is introduced in the literature \cite{26-kataria2017hearing}, which shows an accuracy of about \(54\,cm\) by using binaural amplitude and phase difference as model inputs. 
By comparison, the proposed method significantly improves the accuracy of the sound source location.

\begin{table}[htb]
\renewcommand\arraystretch{1.25}
\centering
\caption{Statistical results of sound source distance estimation methods using the measurement data}
\label{dist error}
\setlength{\tabcolsep}{2.5mm}{
\begin{tabular}{cccc}
\toprule[2pt]
 Method & D-TDOA & D-Height & D-DNN \\
 \midrule[1pt]
 \(S_{mean}\,(m)\) & 1.09 & 0.63 & \textbf{0.33} \\
 \(S_{sd}\,(m)\)   & 0.68 & 0.37 & \textbf{0.34} \\
 \bottomrule[2pt]
\end{tabular}}
\end{table}

\subsubsection{Room geometry inference}
\begin{figure}[hbt]
\centering
\includegraphics[width=20pc]{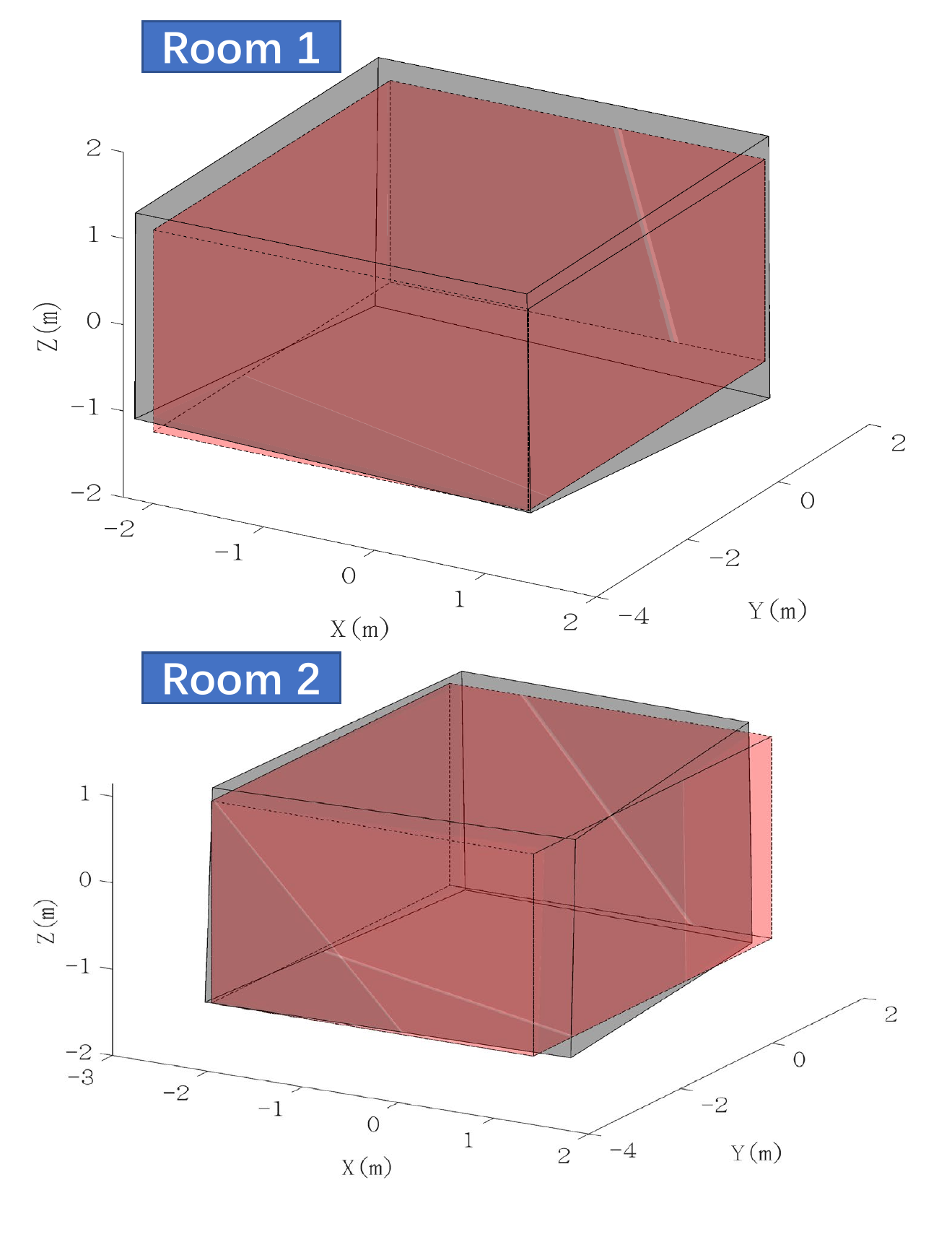}
\caption{Reconstructed results of measured rooms.}
\label{est_rooms}
\end{figure}

Based on the above results, we have reconstructed room 1 and room 2, and the results are shown in Fig.\ref{est_rooms}. The red image represents the actual rectangular room structure, and the gray image represents the reconstructed result. The origin position of the coordinate system is the center of the microphone array. The average distance error and angle error of room1 are \(0.53\,m\) and 4.81°, respectively; The average distance error and angle error of room 2 are \(0.34\,m\) and 4.86 °. Compared with the reconstruction result of room 1, the result of room 2 is more accurate, mainly due to the smaller size of room 2 and the more minor distance error of image sources. In general, the method proposed in this paper can effectively use the single point observation of the sound field to estimate the room boundary in the case of unknown sound source location in the actual environments.

\section{CONCLUSION}
In this paper, we proposed a room geometry blind inference method based on estimating the direct source and first-order reflections in the acoustic enclosure. In this process, the room reverberation information is calculated and used for the localization of the sound source and the estimation of the boundaries with a compact microphone array. Besides, the DNN-based models are designed to realize the time delay estimation and sound source localization with high precision and robustness. The proposed method has two advantages over the conventional room geometry inference techniques. One is that it does not need any prior information about the environment or the measuring of RIR. Apart from that, by using DNN models, the accuracy and integrity of reflection information estimation in the sound field are improved, which is helpful to realize room boundary estimation based on single-point measurement.

Experimental results of both simulations and real measurements verify the effectiveness and accuracy of the proposed techniques compared with the conventional methods under different reverberant environments. According to the measurement data results, the sound source localization error of the proposed DNN model is about 17\%, which is much better than the best results of traditional methods. Besides, the proposed DNN-based models can reduce the distance and angle error of the boundaries estimation results by about 10\% and 25\%, respectively.

\section{ACKNOWLEDGMENT}
This work is supported by the National Key Research
and Development Program (No.2019YFC1408501), the National
Natural Science Foundation of China (No.U1713217, No.61175043, No.61421062), and the High-performance Computing Platform
of Peking University.

\bibliography{room_inference.bbl}
\bibliographystyle{aes2e.bst}

\end{document}